\documentclass[proceedings]{JHEP37}
\usepackage{eurosym}
\usepackage{amssymb}
\usepackage{amsfonts,cite}
\usepackage{amsmath}
\usepackage{graphicx}
\usepackage{multirow}
\usepackage{epsfig}
\usepackage[utf8]{inputenc}
\usepackage{multicol}
\usepackage{graphicx}
\usepackage{wrapfig}
\usepackage{placeins}
\usepackage{slashed}
\usepackage{lipsum}
\usepackage{array}
\usepackage[thinlines]{easytable}
\usepackage{mathrsfs}
\usepackage[export]{adjustbox}
\usepackage{blindtext}

\newbox\mybox

\newcommand{\sign}{\text{sign}}
\newcommand\fverb{\setbox\mybox=\hbox\bgroup\verb}
\newcommand\fverbdo{\egroup\medskip\noindent\fbox{\unhbox\mybox}\ }
\newcommand\fverbit{\egroup\item[\fbox{\unhbox\mybox}]}
\conference{Integrable scattering theory with higher derivative Hamiltonians}
\abstract{We discuss how a standard  scattering theory a of multi-particle theory generalises to systems based on Hamiltonians that involve higher-order derivatives in their quantum mechanical formulation. As concrete examples, we consider Hamiltonian systems built from higher-order charges of Calogero and Calogero-Moser systems. Exploiting the integrability of these systems, we compute the classical phase shifts and briefly comment on the quantum versions of these types of theories.}

\title{Integrable scattering theory with higher derivative Hamiltonians }
\author{Andreas Fring and Bethan Turner\\
 Department of Mathematics, City, University of London, Northampton Square,\\ London EC1V 0HB, UK \\
a.fring@city.ac.uk, bethan.turner.2@city.ac.uk}

\input{tcilatex}
\begin{document}

\section{Introduction}
The vast majority of Hamiltonians describing physical systems contain kinetic terms that are quadratic in momenta in their classical versions, and have second-order derivatives in their quantum mechanical and quantum field theoretical formulations. In general, higher order derivative descriptions are discarded as classical version contain singularities that can be reached in finite time 
and quantum versions of these type of theories inevitable introduce so-called ghosts states that possess negative norms, that lead to a collapse and/or a violation of unitarity \cite{raidal2017quantisation}. Despite these severe deficiencies higher derivative quantum field theoretical models are known to possess the very appealing feature of being renormalizable and are therefore frequently considered as promising candidates for theories of everything (TOE) that include gravity besides all the other known fundamental forces. In that context the theories in the conventional (3+1)-dimensional space need to be embedded into a higher dimensional space, which in turn calls for higher derivative terms. Over the years higher order derivative theories have also been considered sporadically in other areas. They have been proposed as a resolution of the cosmological singularity problem  \cite{biswas2010towards} with some of their black holes solutions constructed \cite{mignemi1992black}. In some quantum field theories their BRST symmetries have been identified \cite{rivelles2003triviality,Kap1}, they were explored in a massless particle description of bosons and fermions \cite{Mpl} and also some supersymmetric versions have been investigated \cite{dine1997comments}. Classical and quantum stability properties of higher derivative dynamics were investigated in \cite{Sugg1,Sugg2,Sugg3,Sugg4}. As discussed in \cite{smilga2005benign,smilga2021exactly,Smilga6,smilga2021benign}, the undesired features of singularities and ghost states may be dealt with in individual cases with appropriate techniques when they are of benign type, i.e. when in their classical versions the unavoidable singularities can not be reached in finite time.  

While there are proposals to deal with some of the conceptual issues in particular cases, the selection of appropriate models has been rather ad hoc, with the Pais-Uhlenbeck oscillator \cite{pais1950field} being the most popular toy model. It remains unclear which type or class of higher derivative theories might be most appropriate and which models contain benign and which contain malevolent ghost states. Recently Smilga  \cite{smilga2021exactly} suggested that Hamiltonians build from higher charges of integrable systems could be a suitable class of benign systems. Indeed, starting at first with classical theories that have kinetic terms with higher order momenta build from charges of affine Toda lattice theories the present authors \cite{bethanAF} identified under which circumstance those theories converge or diverge in phase space. 

Motivated by the reasons mentioned above, and the success obtained by taking integrable systems as suitable candidates for higher derivative theories we continue here the investigation of these models. In particular, we address the new question of how one might consistently formulate a scattering theory. We will answer the question in the context of integrable systems, taking Calogero-Moser-Sutherland (CMS)-models \cite{Cal1,Cal2,Mo,Suth3,Suth4} as suitable candidates to study.

Thus, our general starting point is a system in the form of standard Hamiltonians of CMS type
\begin{equation}
	H= \frac{1}{2}\sum_{i=1}^{\ell}  p_i^2+ 
	\sum_{\alpha \in \Delta_{\bf{g}} }     c_\alpha V\left( \alpha \cdot q \right) .  \label{Calo1H}
\end{equation}
The sum in the potential extends over all roots $\alpha$ in the root space $\Delta_{\bf{g}} \in \mathbb{R}^\ell $ associated to the Lie algebra $\bf{g}$,  $c_{\alpha}$ are real coupling constants, $p=(p_1,\ldots, p_\ell)$ and $q=(q_1,\ldots, q_\ell)$ are the $\ell$ momenta and coordinates of the system, respectively. Requiring integrability, the potential function $V(x)$ may be any function that factorises and satisfies the following functional relation \cite{Func}
\begin{equation}
        V(x)= - f(x)f(-x), \qquad \text{with} \quad  f(x+y) = \frac{f(x)f'(y) - f'(x) f(y)}{V(x)-V(y)} . \label{funcf}
\end{equation}
Solutions to these equations are $f(x)=1/x$, $f(x)=1/\sin(x)$, $f(x)=1/\sinh(x)$ and $f(x)=1/sn(x)$, giving rise to the  Calogero-Moser-Sutherland potentials \cite{Func} $V(x)=1/x^2$, $V(x)=1/\sin^2(x)$, $V(x)=1/\sinh^2(x)$, $V(x)=1/sn^2(x)$, respectively. Here we will focus our attention on the theories that allow for scattering states, i.e. $V(x)=1/x^2$ and $V(x)=1/\sinh^2(x)$.

Our manuscript is organised as follows: In section 2 we make some general comments on how higher order charges can be constructed from Lax pairs corresponding to the higher order flows in the CMS models. In section 3, we formulate the scattering theory for higher charge Hamiltonian theories in the context of integrable systems. In particular, we derive explicit formulas for the two particle phase shifts. In section 4 and 5 we will discuss in detail the higher order scattering theories of Calogero and Calogero-Moser systems associated to the $A_2$ and $A_6$ Lie algebras, respectively. In section 6 we comment on the quantum mechanical scattering theory. Our conclusions are stated in section 7.

\section{Conserved charges in $A_n$-Calogero-Moser-Sutherland models}
The cornerstone of our construction are the higher charges of integrable multi-particle systems. A standard way to construct them is in the form of their Lax pair representation, which implies their classical Liouville integrability. For CMS-models different versions of Lax pairs can be found in the literature for specific algebras,  concrete choices of the representations of the roots and functions $f(x)$, but also in more generic terms see e.g. \cite{sasaki2006explicit}.

We start with the $A_2$-CMS theory with a representation independent formulation for generic functions $f(x)$ subject to the equations (\ref{funcf}). A possible Lax pair is given as
\begin{eqnarray}
	L &=& \left(
	\begin{array}{ccc}
		p_1 & i \sqrt{g} f\left(\alpha _1\cdot q\right) & i \sqrt{g} f\left(\alpha _3\cdot q\right) \\
		-i \sqrt{g} f\left(\alpha _1\cdot q\right) & p_2 & i \sqrt{g} f\left(\alpha _2\cdot q\right) \\
		-i \sqrt{g} f\left(\alpha _3\cdot q\right) & -i \sqrt{g} f\left(\alpha _2\cdot q\right) & p_3 \\
	\end{array}
	\right),   \label{Lop} \\
	 M &=& i \sqrt{g} \left(
	 \begin{array}{ccc}
	 	\frac{f\left(\alpha _3\cdot q\right) f'\left(\alpha _2\cdot q\right)+f\left(\alpha _2\cdot q\right) f'\left(\alpha _3\cdot q\right)}{f\left(\alpha _1\cdot
	 		q\right)} & f'\left(\alpha _1\cdot q\right) & f'\left(\alpha _3\cdot q\right) \\
	 	f'\left(\alpha _1\cdot q\right) & 0 & f'\left(\alpha _2\cdot q\right) \\
	 	f'\left(\alpha _3\cdot q\right) & f'\left(\alpha _2\cdot q\right) & \frac{f\left(\alpha _3\cdot q\right) f'\left(\alpha _1\cdot q\right)+f\left(\alpha _1\cdot
	 		q\right) f'\left(\alpha _3\cdot q\right)}{f\left(\alpha _2\cdot q\right)} \\
	 \end{array}
	 \right), \,\, \quad
\end{eqnarray}
so that the Lax equation $\dot{L} = [L,M]$, \cite{Lax}, becomes equivalent to the equations of motion resulting from the Hamilton $H$ in equation (\ref{Calo1H}) 
\begin{equation} 
 \dot{x}_i = p_i, \qquad	\dot{p}_i = -2 g \sum_{j=1}^3 \alpha_j^i f\left(\alpha_j \cdot q\right) f'\left(\alpha _j\cdot q\right), \quad i=1,2,3. \label{eqm1} 
\end{equation}
The conserved charges are then calculated in the standard fashion from the trace of products of the $L$-operator
\begin{eqnarray} 
	Q_1 &=& \text{tr}(L) =p_1+p_2+p_3,\\
	Q_2 &=& \frac{1}{2} \text{tr}(L^2) =\frac{1}{2} \left( p_1^2+p_2^2+p_3^2  \right)+
	  g \left[f^2\left(\alpha _1\cdot q\right)+f^2\left(\alpha _2\cdot q\right)+f^2\left(\alpha _3\cdot
	q\right)\right], \label{Q2charge}\\
	Q_3 &=& \frac{1}{3} \text{tr}(L^3)=  \frac{1}{3} \left( p_1^3+p_2^3+p_3^3  \right)+
	g \left[ p_{12}f^2\left(\alpha _1\cdot q\right)+ p_{23}f^2\left(\alpha _2\cdot q\right)+ p_{13}f^2\left(\alpha _3\cdot
	q\right)\right]\!, \quad\,\, \, \label{Q3charge} \\
	Q_4 &=& \frac{1}{4}\text{tr}(L^4) = \frac{Q_1^4}{24}-\frac{Q_1^2 Q_2}{2}+Q_1 Q_3+\frac{Q_2^2}{2},\\
	Q_5 &=& \frac{1}{5}\text{tr}(L^5)= \frac{Q_1^5}{80}-\frac{Q_1^3 Q_2}{12}-\frac{Q_1 Q_2^2}{4}+\frac{Q_1 Q_4}{2}+Q_2 Q_3.
\end{eqnarray}
We used the abbreviation $p_{ij}:= p_i+p_j$. As indicated for the first examples, the charges $Q_i$ with $i>3$ can be build from combinations of the first three independent ones. Identifying $Q_2$ with the Hamiltonian in the standard way, the $Q_2$-flow generated by Hamilton's equations $\dot{q}_i = \partial Q_2 / \partial p_i$, $\dot{p}_i = -\partial Q_2 / \partial q_i$  yields the same equations of motion as the Lax pair, i.e. (\ref{eqm1}). The mutual Poisson brackets of the charges vanish
\begin{equation}
	\left\{ Q_i, Q_j  \right\}:= \sum_{k=1}^3\frac{\partial Q_i}{\partial q_k}  \frac{\partial Q_j}{\partial p_k} -\frac{\partial Q_i}{\partial p_k}  \frac{\partial Q_j}{\partial q_k} =0, \qquad \text{for} \,i,j = 1,2,3,
\end{equation}
i.e. they are in involution. Furthermore, we find formally the same Poisson bracket relations between the centre-of-mass coordinate $\chi=q_1+q_2+q_3 $ and the conserved charges as in the case of the affine Toda lattice theories \cite{bethanAF}, that is
\begin{equation}
  \left\{\chi, Q_{n+1}  \right\}= n Q_n, \qquad n =1,2, \ldots.     \label{chiQ}
\end{equation}

Interpreting now instead of $Q_2$ the charge $Q_3$ as the Hamiltonian, i.e. considering the $Q_3$-flow rather than the $Q_2$-flow, Hamilton's equations of motion $\dot{q}_i = \partial Q_3 / \partial p_i$, $\dot{p}_i = -\partial Q_3 / \partial q_i$ become
\begin{eqnarray} 
	\dot{q}_i &=&p_i^2+g \left[f^2\left(\alpha _{i-1}\cdot q\right)+f^2\left(\alpha _i\cdot q\right)\right],  \qquad i=1,2,3, \,\mod 3 \label{Q3H1} \\	
	\dot{p}_i &=& -2 g \sum _{j=1}^3 \alpha _j^i \left(p_j+p_{j+1}\right)  f\left(\alpha _j\cdot q\right) f'\left(\alpha _j\cdot q\right).  \label{eqmq31} 
\end{eqnarray}
Also these equation of motion may be associated to a Lax pair. We present this here for our two main theories of interest. As we do not expect to find new charges for this system we keep the $L$-operator as in (\ref{Lop}), and solely modify the M-operator to 
\begin{equation}
	 M = i \sqrt{g} \left(
	\begin{array}{ccc}
		m_1 & f'\left(\alpha _1\cdot q\right) (p_1 +p_2) -Z_3& f'\left(\alpha _3\cdot q\right) (p_1 +p_3) +Z_2\\
		f'\left(\alpha _1\cdot q\right) (p_1 +p_2) +Z_3& m_2  & f'\left(\alpha _2\cdot q\right)(p_2 +p_3) -Z_1 \\
		f'\left(\alpha _3\cdot q\right)(p_1 +p_3) -Z_2 &  f'\left(\alpha _2\cdot q\right)(p_2 +p_3) + Z_1 & m_3 \\
	\end{array}
	\right),
\end{equation}	
where for the Calogero model when $f(x)=1/x$ we have
\begin{equation} 
	m_i = -i   \sqrt{g}   \! \sum_{\substack{j=1,2,3 \\  j \neq i} }    \!\!\!    h(q_i-q_j) (p_i+ p_j),  
	\quad \text{and} \quad
	Z_i=Z=   g f(q_1 -q_2) f(q_1 -q_3)  f(q_2 -q_3), 
\end{equation}
and for the Calogero-Moser model when $f(x)=1/ \sinh(x)$ we need
\begin{eqnarray} 
	m_1 &=&m_2+i \sqrt{g} \left\{     \frac{h^2(q_1 -q_3)}{f^2(q_1 -q_3)} p_1 -    \frac{h^2(q_2 -q_3)}{f^2(q_2 -q_3)} p_2   
	+    \left[   f^2(q_1 -q_3)  -f^2(q_2 -q_3)       \right]  p_3 \right\} ,  \qquad\\
	m_3 &=&m_2+i \sqrt{g} \left\{     \frac{h^2(q_1 -q_3)}{f^2(q_1 -q_3)} p_3 -    \frac{h^2(q_1 -q_2)}{f^2(q_1 -q_2)} p_2   
+    \left[   f^2(q_1 -q_2)  -f^2(q_1 -q_3)       \right]  p_1 \right\} , \\
	Z_i &=&   Z  \prod_{\substack{j=1,2,3 \\  j \neq i}} \cosh(q_i-q_j) .
\end{eqnarray}
 Using the different $M$-operators in the Lax pair is then equivalent to the equations of motion or the $Q_2$-flow (\ref{eqm1}) and the $Q_3$-flow (\ref{Q3H1}), (\ref{eqmq31}), respectively. 
 
 The generalization of the Lax pairs to the $A_n$-higher-charge theories is straightforward. We will not report this here, but appeal below to the fact that this is possible.
 
 \section{Integrable scattering in higher charge Hamiltonian theories}
 When the function $f(x)$ is chosen in such a way that we obtain a scattering theory, that is for the Calogero and Calogero-Moser system, we will encounter classical phase shifts in the scattering process by comparing the asymptotes of the in-state with those of the out-state. In general, the classical phase shifts $\delta_i$, are defined as the asymptotic difference in the coordinates when comparing the free particle motion with an interacting particle of the same momentum. For integrable systems there exists a second possibility leading to the same overall shift by adding the consecutive two-particle phase shifts with the appropriate signs. 
 
 Next, we discuss how the standard scattering theory generalises to a scattering theory based on higher charge Hamiltonians. The potentials we consider here are impenetrable and repulsive, preventing any particle from catching up with another during the scattering process. Consequently, the spatial ordering of the particles on the line remains unchanged at all moments in time.  In a $H=Q_m$-theories for $m>2$ we no longer have pure potentials, that is terms depending only on the coordinates. Instead, the potentials are multiplied by momenta, which are, however, finite. Thus, it remains still impossible to cross the singularities in the potentials to re-arrange for a new ordering. This means the spacial ordering remains fixed for all of these theories too. For definiteness, we select her for this argument the ordering 
 \begin{equation}
 	q_i(t) > q_{i+1}(t),   \qquad  \forall t, \, i=1,2, \ldots, n+1 ,  \label{qordering}
 \end{equation}
 with $n$ being the rank of the $A_n$ Lie algebra in the concrete example to be considered below.
 
 Next, we consider the behaviour of the momenta in the asymptotic in and out-states, $p_i^-$ and $p_i^+$, respectively. We exploit the fact that the theories are classically integrable and posses a Lax pair formulation. As we have seen above for higher charge Hamiltonian systems this can also be achieved by keeping the  $L$-operator unchanged and modifying the  $M$-operator appropriately. We recall now the well-known fact that the eigenvalues of the $L$-operator are preserved, due to the form of the solution to the Lax equation $L(t) = \exp(- M t ) L(0)   \exp(M t )$. In our concrete case we have $\lim_{t \rightarrow \pm \infty} L_{ij} = p_i^\pm \delta_{ij}$ for all $H=Q_m$-theories. Thus the two sets of asymptotic momenta $\{p_1^-, p_2^-, \ldots, p_{n+1}^-  \}$ and $\{p_1^+, p_2^+, \ldots, p_{n+1}^+  \}$, which are the eigenvalues of $L$, must be identical. The only possible change can be a re-ordering when associating them to the coordinates, which as we argued remain unchanged throughout the entire scattering process. We stress, that since this reasoning does not depend on the form of the $M$-operator, but solely on the fact that the equations of motion can be written equivalently as a Lax pair, it also holds when $H=Q_m$ with $m>2$.  The spacial ordering is governed by the velocities with $\dot{q}_i^- <  \dot{q}_{i+1}^- $ in the in-state and $\dot{q}_i^+ >  \dot{q}_{i+1}^+ $ in the out-state. Given our choice (\ref{qordering}), any other ordering would not be asymptotic. Given the general form of the charges $Q_m$, the asymptotic velocities behave as $\lim_{t \rightarrow \pm \infty} \dot{q}_i \sim (p_i^\pm)^{m-1} $. Thus, for a $H=Q_m$-theory we have the orderings  
 \begin{equation}
 	\left( p_i^- \right)^m <  \left( p_{i+1}^- \right) ^m ,  \qquad  \text{and}  \qquad \left( p_i^+\right)^m >  \left( p_{i+1}^+ \right)^m,  \qquad   \, i=1,2, \ldots, n+1 .  \label{pordering}
 \end{equation}
 Evidently this distinguishes a standard Hamiltonian theory, $H=Q_2$, from the higher charge Hamiltonian theories, where in the former of the momenta govern the ordering, whereas in the latter we have to take higher powers into account.

 Next we discuss the classical phase shifts for the two-particle scattering between particle $i$ and $j$. For the standard Hamiltonian theory they have been computed for the $A_n$-Calogero and the Calogero-Moser models, see  \cite{arutyunov19,kulishfact}. Translating the results into our notations and conventions they read
  \begin{equation}
  \Delta_{ij}^C =0, \qquad  \text{and} \qquad	\Delta_{ij}^{CM} =   \frac{1}{2}  \ln\left[  1 +  \frac{4 g }{ (p_i^- -p_j^- )^2 }   \right],  \qquad   \, i,j=1,2, \ldots, n+1 , \label{deltashift1}
  \end{equation}
where $p_i^-,p_j^-$ are the asymptotic momenta of the respective particles mentioned above for $t \rightarrow -\infty$. We will now show that one obtains the same two-particle phase shifts when taking higher charges as Hamiltonians, but the overall shift differs in a multi-particle scattering event. In general, for a $H=Q_m$-theory with repulsive potentials the overall phase shift of particle $i$ is
\begin{equation}
	\delta_i ^m=  \sum_j   \sign\left[  (p_i^-)^{m-1} - (p_j^-)^{m-1}    \right]  \Delta_{ij},    \label{sumshiftgen1}
\end{equation}   
where the sum over $j$ extends over all particles $i$ scatters with and $\sign (x)$ denotes the standard signum function, i.e. $\sign (x)=1$ for $x>0$ and $\sign (x)=-1$ for $x<0$.  We may think of $\sign\left[  (p_i^-)^{m-1} - (p_j^-)^{m-1}    \right]$ as $\sign\left( q_i - q_j  \right)$ so that this formula would be universal with no explicit reference made to the charge that corresponds to the Hamiltonian. However, the ordering in the coordinates for  an  $H=Q_m$-theory is governed by the asymptotic velocities as $\lim_{t \rightarrow \pm \infty} \dot{q}_i \sim (p_i^\pm)^{m-1} $ rather than just by the momenta.

We now show how (\ref{sumshiftgen1}) is obtained in general, by first deriving the two-particle phase shift (\ref{deltashift1}) from the two-particle theory for higher charge Hamiltonians along the lines of \cite{arutyunov19,kulishfact}.  The conserved charges in involution are in this case
\begin{eqnarray}
	Q_1&=& p_1 + p_2 ,  \\
	Q_2&=& \frac{1}{2} \left( p_1^2 + p_2^2  \right)  + g  V(q), \\
	Q_3&=& \frac{1}{3} \left( p_1^3 + p_2^3  \right)  + g  \left( p_1 + p_2 \right) V(q) = Q_1 Q_2 -\frac{1}{6}  Q_1^3 ,\\
	   &\vdots&       \notag \\
	   Q_n&=&\sum_{l=0}^{N}c_lQ_1^{a_l}Q_2^{\frac{1}{2}(n-a_l)},  \quad
	  \left\{  \begin{array}{ l c l l }
	   	     N &=& \frac{1}{2}(n-1) ,   a_l=2l+1   &  \quad \text{for}   \,  n \,  \text{odd}  \\ 
	   	     N &=& \frac{1}{2}n, a_l=2l                &  \quad  \text{for}  \,  n  \, \text{even}  
	   	\end{array}    \right.   ,
\end{eqnarray} 
where for definiteness we chose $q=q_2-q_1>0$ and $c_l \in \mathbb{R}$ are arbitrary constants at this point. Since these charges are conserved they imply the equalities
\begin{eqnarray}
	p_1 + p_2&=& p_1 ^- + p_2^- ,  \label{123} \\
	\frac{1}{2} \left( p_1^2 + p_2^2  \right)  + g  V(q) &=& \frac{1}{2} \left[ (p_1^-)^2 + (p_2^-)^2  \right] ,  \label{234}\\
	&\vdots&  \notag  \\  
		Q_n &=& \frac{1}{n} \left[ (p_1^-)^n + (p_2^-)^n  \right] ,  \label{567}
\end{eqnarray}
where on the right hand side of (\ref{234}), (\ref{567}) we have taken into account that $\lim_{t \rightarrow - \infty} V(q)=0$, i.e. all particles move infinitely far away from each other in the in and out-states and fixed the constants $c_l $ to obtain the right hand side of (\ref{567}). Solving (\ref{123}), (\ref{234}) or  in general (\ref{123}), (\ref{567}) we obtain in each case
\begin{equation}
  \left.  \begin{tabular}{ l l }
 $ p_1(t)$ \!\!\! &$ \!\!\! \!\! =\frac{1}{2}    \left[  p_1^- + p_2^-   \pm \sqrt{  k^2  - 4 g V(q)     }       \right] $ \\
  $p_2(t)$ \!\!\!  &$\!\!\! \!\! = \frac{1}{2}    \left[  p_1^-  +  p_2^-   \mp \sqrt{  k^2  - 4 g V(q)     }       \right] $
   \end{tabular}  \right\} \quad \text{for } \, t  \,  \substack{< \\ >} \, t_0  .    \label{p12eq}
\end{equation}
We abbreviated here $k:=p_1^- - p_2^- >0$ with the positivity being implied by our ordering in the coordinates. The time at the turning point when $k^2 = 4 g V(q_0)$ is denoted as $t_0$, so that the upper sign refers to the time before the scattering process and the lower sign to the situation afterwards. 

Up to this point, there are no discernible differences among the various scenarios when interpreting different charges as Hamiltonians. However, now we will use the general equation of motions $\dot{q}_i = \partial Q_m / \partial p_i$ rather than (\ref{eqm1}) for the evolution of the coordinate difference 
\begin{equation}
	\dot{q}=  \dot{q}_2 -     \dot{q}_1  =    \mp\frac{1}{2}\sum^{N}_{l=0}c_l(n-a_l)(p_2-p_1)Q_1^{a_l}Q_2^{\frac{1}{2}(n-(a_l+1))}          , \quad \text{for } \, t  \,  \substack{< \\ >} \, t_0.
\end{equation}
Notice that the explicit $p$ and $q$ is reduced to the factor $(p_2-p_1)$. Within the charges this dependence can be replaced by their asymptotic values. Replacing now the momenta using (\ref{p12eq}) we obtain 
\begin{equation}
	\dot{q}=\mp K\sqrt{1-\frac{4gV(q)}{k^2}},   \quad \text{for } \, t  \,  \substack{< \\ >} \, t_0,
	\label{qdot2}
\end{equation} 
with constant $K:=\frac{1}{2}\sum^{N}_{l=0}c_l(n-a_l)Q_1^{a_l}Q_2^{\frac{1}{2}(n-(a_l+1))}$.
The equations are easily integrated out by separation of variables. For the Calogero model and Calogero-Moser model we obtain
\begin{eqnarray}
	t&=&  \mp \frac{1}{  K} \sqrt{ q^2- \frac{4 g}{k^2}} + a_C^\mp, \qquad   \qquad   \qquad   \qquad   \qquad \,\,\,\, \text{for }  V(q)=\frac{1}{q^2} ,\\
	t &=&  \mp \frac{1}{  K} \ln \left[\cosh q+ \frac{1}{k} \sqrt{ k^2 \sinh^2q - 4 g}  \right]+ a_{CM}^\mp, \quad \text{for }  V(q)=\frac{1}{\sinh^2q},  \qquad
\end{eqnarray}
respectively. Where $a_C^\mp$ and $ a_{CM}^\mp$ are integration constants. A relation between the constants for  $t    <  t_0$ and $t    >  t_0$ is obtained by demanding that both solutions coincide at the turning point ($t_0$,$q_0$). This requirement yields
\begin{equation}
a_C^- = a_C^+, \qquad \text{and} \qquad a_{CM}^- =  \frac{2 \ln \left( \cosh q_0 \right) }{ K} +    a_{CM}^+ =\frac{1 }{ K }  \ln \left(  1 + \frac{4 g}{k^2}    \right)+    a_{CM}^+   . \label{constmatch}
\end{equation}
Further constraints on the constants arise from matching with the asymptotic behaviour arising directly from the equations of motion
\begin{equation}
q(t )= K t  +  q_2^\mp -  q_1^\mp, \quad \text{for } \, t  \,  \substack{< \\ >} \, t_0.
\end{equation}
This leads to the identifications
\begin{equation}
    a_C^- =  \frac{q_2^- - q_1^-}{K}=a_C^+ =  \frac{q_1^+  -q_2^+}{K},  \qquad  a_{CM}^- = \frac{q_2^- - q_1^-}{K}= a_{CM}^+ =  \frac{q_1^+  -q_2^+}{K}, 
\end{equation}
which when combined with (\ref{constmatch}) and the fact that  $\delta_1+\delta_2=0$ leads to
\begin{eqnarray}
	\delta_1^C&=& q_1^+ -q_2^- =q_2^+ -q_1^- = 	\delta_2^C, \qquad  \Rightarrow  \qquad \delta_1^C=\delta_2^C=0, \\
	\delta_2^{CM}&=&  \ln \left(  1 + \frac{4 g}{k^2}    \right) + \delta_1^{CM} \qquad  \qquad \,\, \Rightarrow \qquad  \delta_2^{CM}=-\delta_1^{CM}=\frac{1}{2}  \ln \left(  1 + \frac{4 g}{k^2}   \right)   . \,\,\,\,\,\,
 \end{eqnarray}
Notice that the sum in (\ref{sumshiftgen1}) only includes particle $2$. Next, we illustrate the above general statements with concrete examples.

\section{A higher charge $A_2$-Calogero model}
\begin{figure}[h]
	\centering         
	\begin{minipage}[b]{0.52\textwidth}           \includegraphics[width=\textwidth]{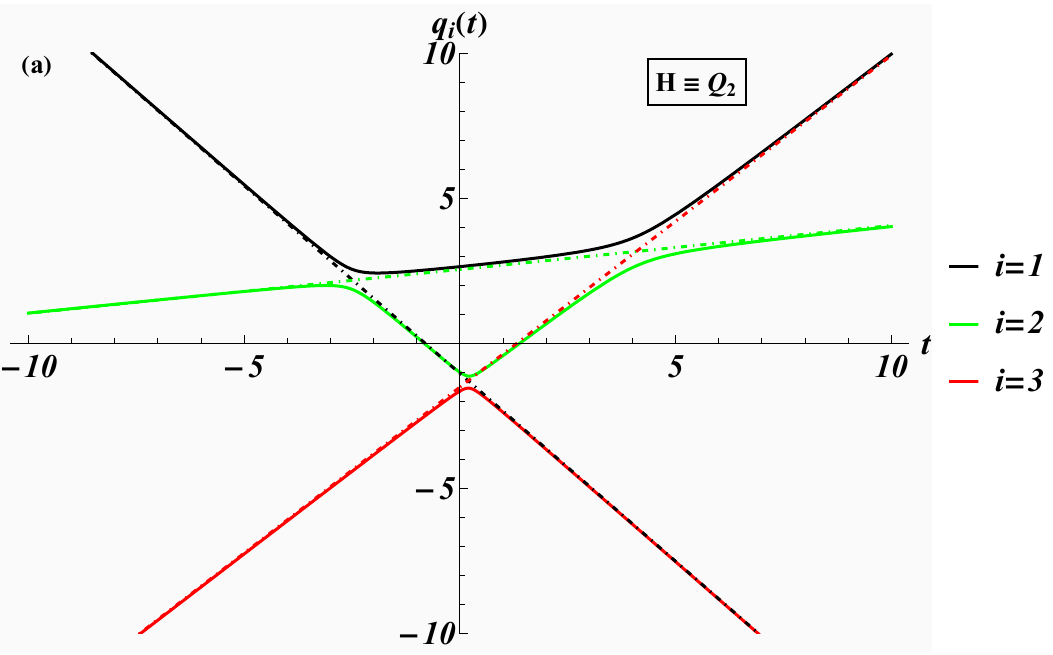}
	\end{minipage}   
	\begin{minipage}[b]{0.47\textwidth}           
		\includegraphics[width=\textwidth]{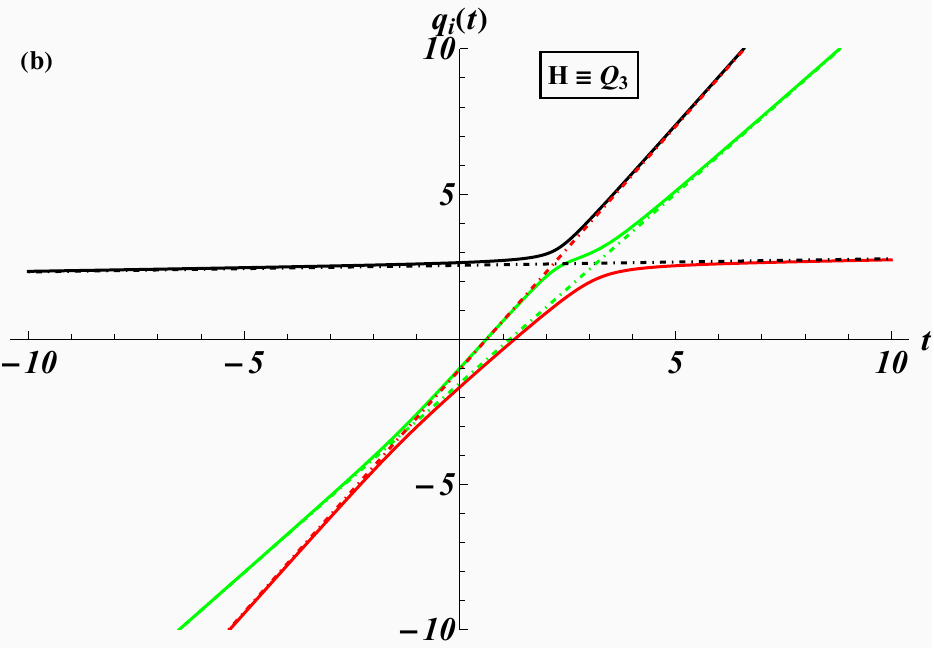}
	\end{minipage}  
	\caption{Coordinates as functions of time for the $A_2$-Calogero model with $Q_2=H$ (panel a) and $Q_3=H$ (panel b) (solid lines) together with their asymptotes (dashed lines). The initial conditions respect $Q_1=\chi=0$ and are taken to $q_1(0)=2.63998$, $q_2(0)=-0.994087$, $q_3(0)=-1.64589$, $p_1(0)=0.152777$, $p_2(0)=-1.00456$, $p_3(0)=0.851788$ corresponding to $E=1.5$, $B=4$, $g=0.25$, $\phi_0 = -3$, $t_0=0$, $R_0=0$, $R_1=0$.  For these choices we have $q_1^a = -1.50968$,  $q_2^a = 2.5458$ and $q_3^a = -1.03612$ for the values of the asymptotes at $t=0$ . The coupling constant is taken to $g=1/4$.} 
	\label{Calo1}
\end{figure}

For the $A_2$-Calogero model with $f(x)=1/x$,  i.e. $V(x)=1/x^2$, the coupling constants set equal $c_\alpha = g$ and the simple roots taken in the standard three dimensional representation $\alpha_1=(1,-1,0)$, $\alpha_2=(0,1,-1)$, $\alpha_3=\alpha_1+\alpha_2=(1,0,-1) $, see e.g. \cite{bourbaki1968groupes}, an explicit analytical solution to the equations of motion (\ref{eqm1}) with $Q_2$ taken as Hamiltonian 
\begin{equation}
	\dot{q}_i = p_i, \qquad 	\dot{p}_i = 2 g \sum_{\substack{k=1,2,3 \\ i \neq k} } \frac{1}{(q_i - q_k)^3}	,
\end{equation}
	was already obtained in \cite{marchiorosol} by separating variables 
\begin{eqnarray} 
q_1(t) &=&  R_0 + R_1 t   +   \frac{1}{\sqrt{6}}  r(t) \cos [\phi (t)]  +   \frac{1}{\sqrt{2}}  r(t) \sin [\phi (t)] ,   \\
q_2(t) &=&  R_0 + R_1 t   +   \frac{1}{\sqrt{6}}  r(t) \cos [\phi (t)]  -   \frac{1}{\sqrt{2}}  r(t) \sin [\phi (t)] ,\\
q_3(t) &=&  R_0 + R_1 t     -   \frac{2}{\sqrt{6}}  r(t) \cos [\phi (t)] .
\end{eqnarray}
The functions and the constant
\begin{eqnarray} 
	\phi(t) &:=&  \frac{1}{3} \arccos \left\{ \kappa  \sin \left[  \arcsin\left(\frac{\cos \left(3 \phi_0\right)}{\kappa }  \right) -3 \arctan\left(  \frac{\sqrt{2} \text{E}
		\left(t-t_0\right)}{B}\right) \right]  \right\} ,\\
			r(t) &:=& \sqrt{\frac{B^2}{\text{E}}+2 \text{E} \left(t-t_0\right)^2}, \qquad
			\kappa := \sqrt{1-\frac{9 g}{2 B^2}},
\end{eqnarray}
depend on the parameters $E,B,R_0,R_1,g,\phi_0,t_0 \in \mathbb{R}$, with $E$ being the energy and $B$ an angular constant of motion. The dependent parameters may be expressed entirely in terms of the initial conditions $q_i(t_0)=:q_i^0$ as
\begin{equation} 
	\phi_0 =\arccos \left(    \frac{q_1^0+q_2^0 - 2 q_3^0}{2 \lambda}  \right), \qquad \text{and} \qquad B =  \lambda \sqrt{ \frac{2}{3} E } ,
\end{equation}
with
\begin{equation} 
	\lambda :=  \sqrt{  \left(q_1^0\right)^2   + \left(q_2^0\right)^2  +\left(q_3^0\right)^2   -   q_1^0 q_2^0   -   q_1^0 q_3^0 -   q_2^0 q_3^0      } ,
\end{equation}
which is useful for the numerical considerations. The asymptotic behaviour for these solutions is found to be 
\begin{eqnarray} 
	&&\lim_{t \rightarrow + \infty} q_1(t) \sim q_1^a + p_{1}^{a} t, \quad
	\lim_{t \rightarrow + \infty} q_2(t) \sim q_2^a + p_{2}^{a} t, \quad
	\lim_{t \rightarrow  + \infty} q_3(t) \sim q_3^a + p_{3}^{a} t, \\
	&&\lim_{t \rightarrow - \infty}   q_1(t) \sim q_3^a + p_{3}^{a} t, \quad
	\lim_{t \rightarrow - \infty} q_2(t) \sim q_2^a + p_{2}^{a} t, \quad
	\lim_{t \rightarrow - \infty} q_3(t) \sim q_1^a + p_{1}^{a} t. 
\end{eqnarray}
We compute the values for the asymptotic momenta $p_i^a$ and the crossing values of the asymptotes with the $y$-axis  $q_i^a$ from 
\begin{eqnarray}
	p_i^a&=&p_i^+=\lim_{t \rightarrow + \infty}  \dot{q}_i(t), \qquad \quad \text{and} \qquad  q_i^a=\lim_{t \rightarrow + \infty} \left[ q_i(t) - p_i^a  t \right],\\
	p_i^a&=&p_{4-i}^-=\lim_{t \rightarrow -\infty}  \dot{q}_{4-i}(t), \qquad \text{and} \qquad  q_i^a=\lim_{t \rightarrow -\infty} \left[ q_{4-i}(t) - p_{4- i}^a t \right].
\end{eqnarray}
Setting $R_0=R_1=0$ in the explicit solutions, we obtain
\begin{eqnarray}
p_{1/2}^{a} &=&  \sqrt{E} \left[ \frac{ 1 }{\sqrt{3}} \cos (\varphi ) \pm     \sin (\varphi )  \right], \quad  p_{3}^{a} = - \sqrt{E}\frac{ 2 }{\sqrt{3}} \cos (\varphi ),     \label{pasym1} \\
q_{1/2}^{a} &=& \frac{B \cos \left(3 \phi _0\right) \left(3 \cos (\varphi )-\sqrt{3} \sin (\varphi
	)\right)}{3 \sqrt{E} \sqrt{1+\frac{9 g}{B^2}+\cos \left(6 \phi _0\right)}}, \quad q_3^a = \frac{2 B \sin (\varphi ) \cos \left(3 \phi _0\right)}{3 \sqrt{E} \sqrt{\frac{9
			g}{B^2}+\cos \left(6 \phi _0\right)+1}} ,   \qquad    \label{qasym1} \\
\varphi  &:=& \frac{1}{3} \arccos \left(\sqrt{\kappa -\cos ^2\left(3 \phi _0\right)}\right).
\end{eqnarray}
In figure \ref{Calo1} panel (a) we depict these solutions for the ordering chosen in (\ref{qordering}), i.e. $q_1>q_2>q_3$, together with their asymptotes. 

\begin{figure}[h]
	\centering         
	\begin{minipage}[b]{0.525\textwidth}           
		\includegraphics[width=\textwidth]{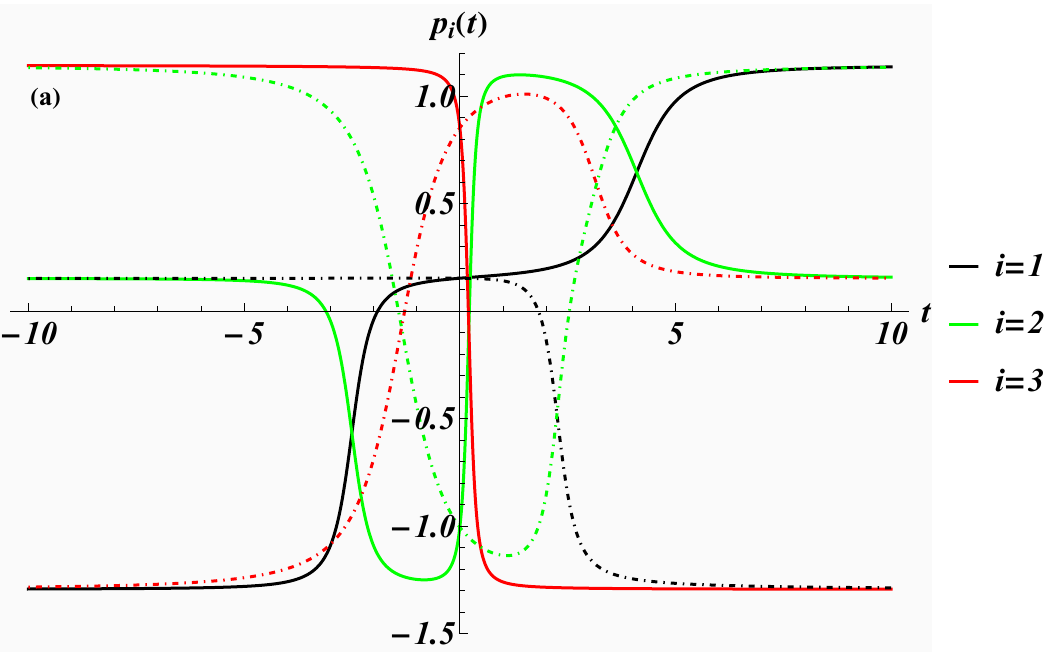}
	\end{minipage}  
	\begin{minipage}[b]{0.455\textwidth}           
		\includegraphics[width=\textwidth]{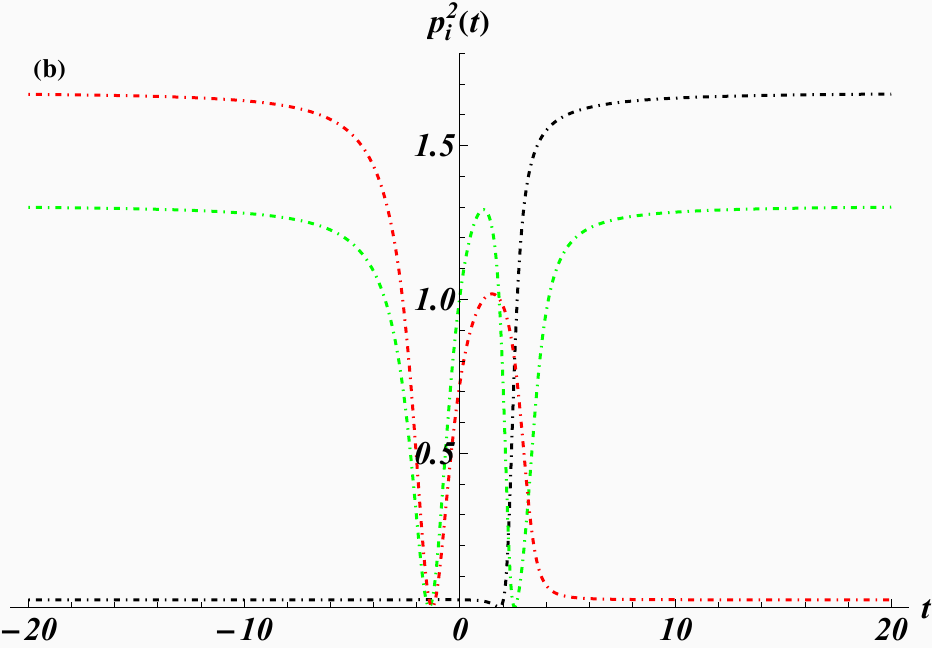}
	\end{minipage}  
	\caption{Momenta as functions of time for the $A_2$-Calogero model  for $Q_2=H$ (solid lines) and $Q_3=H$ (dashed lines) in panel (a) and squared momenta as functions of time in panel (b). The initial conditions and coupling constant are the same as in figure \ref{Calo1}. When $H=Q_2$ the values for the asymptotic momenta are $p_1^-=p_3^+= -1.24871$, $p_2^-= p_2^+=0.15298$,  $p_3^-= p_1^+=1.09573$, which when $H=Q_3$ are re-ordered to $p_3^-=p_1^+= -1.24871$, $p_1^-= p_3^+=0.15298$,  $p_2^-= p_2^+=1.09573$.} 
	\label{Calo2}
\end{figure}

Next we compare these results with the numerical solutions for the system (\ref{Q3H1}), (\ref{eqmq31}) when $Q_3$ is taken to be the Hamiltonian with the same function $f(x)$ and the same representation for the simple roots. In this case the equations of motion are
\begin{equation}
	\dot{q}_i = p_i^2  + g \sum_{\substack{k=1,2,3 \\ i \neq k} } \frac{1}{(q_i - q_k)^2}, 
	\quad \dot{p}_i = 2 g \sum_{\substack{k=1,2,3 \\ i \neq k} } \frac{p_i + p_k}{(q_i - q_k)^3}.   \label{equnofmq3}
\end{equation}
From figures \ref{Calo1} and \ref{Calo2} panel (a) we observe that while the asymptotic behaviour for the coordinates is different in both cases, the asymptotic values for the momenta are identical up to permutations of the particle type. The inequalities for the ordering of the powers of the momenta in (\ref{qordering}) and (\ref{pordering}) are confirmed.

However, as we will see next while the scattering is governed by the ordering of the momenta when $H=Q_2$, it is governed by the ordering of momenta squared when $H=Q_3$.  

Indeed, when $H=Q_2$ we observe that the ordering of the momenta  in the in-state $p_3^- > p_2^- > 0 > p_1^-$ at $t \rightarrow -\infty$ changes for $t \rightarrow \infty$ into 
$p_1^+ < 0 <  p_2^+ <  p_3^+$ in the out-state, i.e. particle 1 undergoes a head on collision with particle 2 and 3, whereas particle 3 scatters with particle 1 by overtaking it due to its larger momentum in the same direction. After the scattering event all particles move away from each other. 

In contrast, when $H=Q_3$ this interpretation does not hold. As we observe in figure 2 panel (a), in this case we have the ordering $p_3^- < 0 < p_1^-  <   p_2^-$ in the in-state, which if the scattering were governed by the momenta would suggest for the given spacial ordering that particle 3 would simple move away to the left without scattering with any other particles and the only scattering process would be between particle 1 and 2 with the latter overtaking the former. Clearly this is not what we observe in figure 2 panel (a). Instead,  we see from the $H=Q_3$-equations of motion (\ref{equnofmq3}) that $\lim_{t \rightarrow \pm \infty} \dot{q}_i(t) \sim  p_i^2$, since $\lim_{t \rightarrow \pm \infty} | q_i(t)  - q_j(t) | \sim  \infty$, so that the squared momenta govern the asymptotic behaviour. This feature is precisely confirmed by figure 2 panel (b), where we observe that the ordering $(p_3^-)^2 >  (p_2^-)^2 > 0 > (p_1^-)^2$ changes into $(p_1^+)^2 >  (p_2^+)^2 > 0 > (p_3^+)^2$, i.e. particle 3 is overtaking particle 1 and 2 whereas particle 2 is overtaking particle 1.

Next we compute the classical phase shifts $\delta_i$, defined as the asymptotic shift in the coordinates when comparing the free particle with a Calogero particle of the same momentum for the two different Hamiltonians. 

When $H=Q_2$ they are computed from the analytical expressions derived above to  
\begin{eqnarray}
	\delta_1 &=&   \lim_{t \rightarrow \infty} \left[  q_1(t) -  q_3(-t)  - 2 p_1^a t   \right] =q_1^a - q_1^a =0 , \\
	\delta_2 &=&   \lim_{t \rightarrow \infty} \left[  q_2(t) -  q_2(-t)  - 2 p_2^a t   \right] =q_2^a - q_2^a =0  ,\\
	\delta_3 &=&  \lim_{t \rightarrow \infty} \left[  q_3(t) -  q_1(-t)  - 2 p_3^a t   \right] =q_3^a - q_3^a = 0  .
\end{eqnarray}
This zero phase shift is indeed what we expect, as the classical phase shift in any two-particle scattering process in the Calogero model is zero, as we have seen.

\noindent When $H=Q_3$ we compute the asymptotic behaviour numerically, obtaining 
\begin{eqnarray} 
	&&\lim_{t \rightarrow + \infty} q_1(t) \sim q_3^a + \left( p_{3}^{a}\right)^2 t, \quad
	\lim_{t \rightarrow + \infty} q_2(t) \sim q_1^a +  \left( p_{1}^{a}\right)^2 t, \quad
	\lim_{t \rightarrow  + \infty} q_3(t) \sim q_2^a +  \left( p_{2}^{a}\right)^2 t, \qquad \quad\\
	&&\lim_{t \rightarrow - \infty}   q_1(t) \sim q_2^a +  \left( p_{2}^{a}\right)^2 t, \quad
	\lim_{t \rightarrow - \infty} q_2(t) \sim q_1^a +  \left( p_{1}^{a}\right)^2 t, \quad
	\lim_{t \rightarrow - \infty} q_3(t) \sim q_3^a +  \left( p_{3}^{a}\right)^2 t. 
\end{eqnarray}
The numerical values for the asymptotics expressions of $q_i$ and $p_i$ are reported in the captions of figures \ref{Calo1} and \ref{Calo2} and coincide precisely with those obtained from the analytical expressions (\ref{pasym1}) and (\ref{qasym1}) for the given initial conditions.

The classical phase shifts are then  
\begin{eqnarray}
	\delta_1 &=&  \lim_{t \rightarrow \infty} \left[  q_1(t) -  q_3(-t)  - 2 \left( p_{2}^{a}\right)^2 t   \right] =q_3^a - q_3^a = 0 , \\
	\delta_2 &=&   \lim_{t \rightarrow \infty} \left[  q_2(t) -  q_2(-t)  - 2 \left( p_{1}^{a}\right)^2 t   \right] =q_1^a - q_1^a =0  ,\\
	\delta_3 &=&   \lim_{t \rightarrow \infty} \left[  q_3(t) -  q_1(-t)  - 2 \left( p_{2}^{a}\right)^2 t   \right] =q_2^a - q_2^a =0  .
\end{eqnarray}
Thus, while the asymptotic behaviour for the coordinates is different when taking either $Q_2$ or $Q_3$ as the Hamiltonian, the overall classical phase shifts are identical to zero in both cases. 

We conclude this section with a brief remark on the sensitivity with regard to the chosen initial conditions. For the coordinates the choice was made to implement the centre-of-mass condition. For the momenta we took the $p_i(0)$ such that $Q_1 =0$. From the Poisson bracket relations (\ref{chiQ}) follows that different choices with $Q_1 \neq 0$ will simply add this value to all momenta for the standard $H=Q_2$-theory and for a general $H=Q_m$-theory it will add to each velocity an overall constant $ m Q_m$, as this is non-vanishing. One might eliminate this shift by different choices of the initial momenta. However, for the sake of comparison, we opted to use the same initial conditions in all cases. Thus, unlike as in bounded motions \cite{bethanAF}, where one finds a strong sensitivity in regard to the initial conditions with non-vanishing right hand sides in (\ref{chiQ}) leading to divergencies, for the scattering theory the overall shifts do not produce any qualitative change in behaviour. We will use these type of initial conditions throughout the manuscript.

Next we consider a model that contains particles that scatter with non-zero phase shifts.

 \section{Higher charge $A_n$-Calogero-Moser models}
 We compare the results of the previous subsection now with a similar calculation carried out for the $A_2$-Calogero-Moser model with $f(x)=1/\sinh(x)$,  i.e. $V(x)=1/\sinh^2(x)$ and the representation of the simple roots taken to be the same as in the previous section. In this case the equations of motion (\ref{eqm1}) with $Q_2$ taken as Hamiltonian become
 \begin{equation}
 	\dot{q}_i = p_i, \qquad 	\dot{p}_i = 2 g \sum_{\substack{k=1,2,3 \\ i \neq k} } \frac{\coth(q_i - q_k)}{\sinh^2(q_i - q_k)}	.
 \end{equation}
  Taking instead $Q_3$ to be the Hamiltonian equations of motion become
 \begin{equation}
 	\dot{q}_i = p_i^2  + g \sum_{\substack{k=1,2,3 \\ i \neq k} } \frac{1}{\sinh^2(q_i - q_k)}, 
 	\quad \dot{p}_i = 2 g \sum_{\substack{k=1,2,3 \\ i \neq k} } \frac{\coth(q_i - q_k)}{\sinh^2(q_i - q_k)}\left(p_i + p_k\right).
 \end{equation}
 Some numerical solutions for $q_i(t)$ are depicted in figure \ref{Calo4}. We do not present the solutions for the momenta as they are very similar to those of the Calogero model seen in figure \ref{Calo2}.
 
 \subsection{$A_2$- higher charge Calogero-Moser model}
\begin{figure}[h]
	\centering         
	\begin{minipage}[b]{0.53\textwidth}           \includegraphics[width=\textwidth]{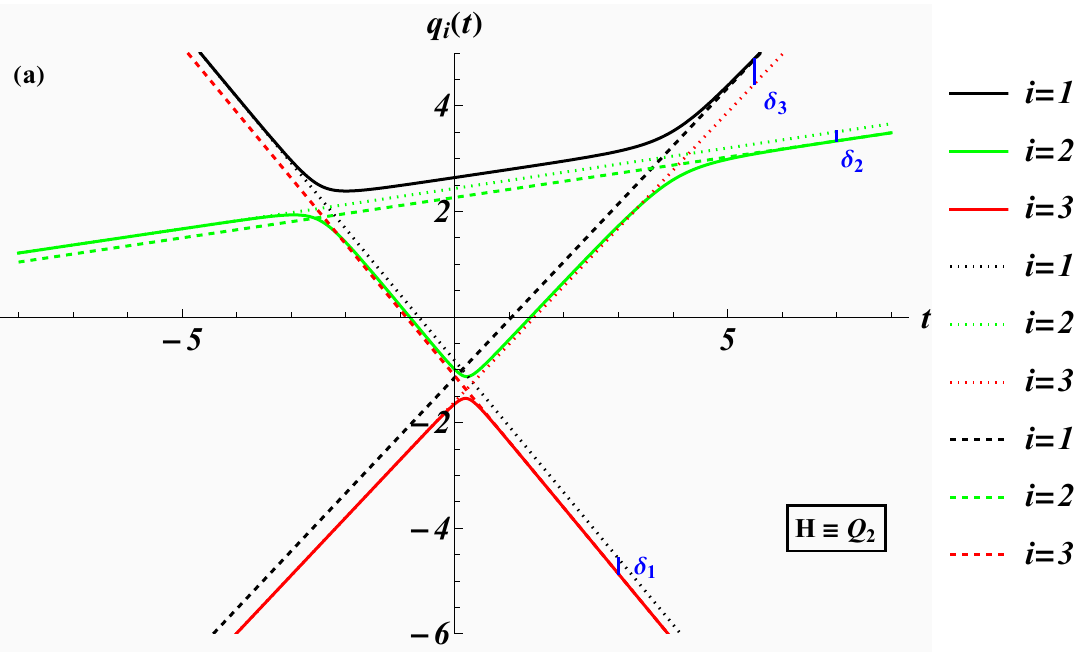}
	\end{minipage}   
	\begin{minipage}[b]{0.46\textwidth}           
		\includegraphics[width=\textwidth]{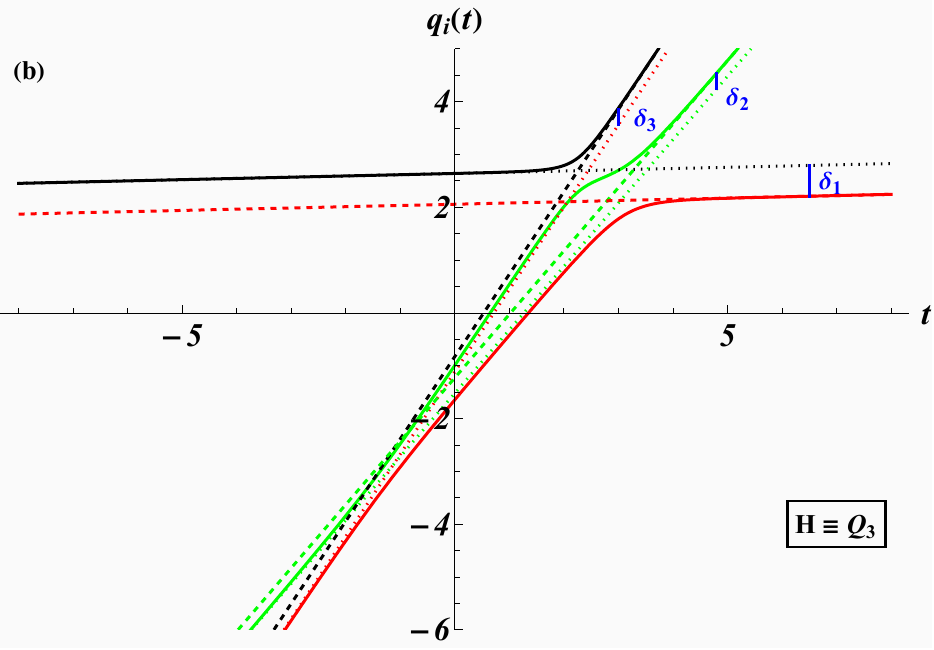}
	\end{minipage}  
	\caption{Coordinates as functions of time for the $A_2$-Calogero-Moser model with $Q_2=H$ (panel a) and $Q_3=H$ (panel b) (solid lines) together with their in-state (dotted lines) and out-state asymptotes (dashed lines). The initial conditions are the same as in figure \ref{Calo1}.  For the asymptotes at $t=0$ we obtain for $H=Q_2$ the values $(q_1^a)^+  =-1.14586$,  $(q_2^a)^+  =2.26276$, $(q_3^a)^+  =-1.11690$, $(q_1^a)^-  =-0.82760$,  $(q_2^a)^- =2.43396$, $(q_3^a)^-  =-1.60636$ and for $H=Q_3$ we have $(q_1^a)^+  =-0.82760$,  $(q_2^a)^+  =-1.22944$, $(q_3^a)^+  =2.05704$, $(q_1^a)^-  =2.63968$,  $(q_2^a)^- =-1.52278$, $(q_3^a)^-  =-1.1169$. } 
	\label{Calo4}
\end{figure}

 For the Calogero-Moser systems the two-particle phase shift $\Delta_{ij}$ and the overall classical phase shifts $\delta_i$ can be computed from (\ref{deltashift1}) and (\ref{sumshiftgen1}), respectively.  Starting with the $H=Q_2$-theory, for our ordering $q_3>q_2>q_1$, implying $p_3^- > p_2^- > p_1^-$, the sign in overall phase shift for particle $i$ are determined from $ \sign (p_i^- - p_j^-) $. We confirm the general formulas with the numerical values obtained from solution depicted in figure \ref{Calo4}
 \begin{eqnarray}
 	\delta_1&=& -\Delta_{12} -\Delta_{13}= (q_3^a)^+ - (q_1^a)^- =  -0.28929 , \\
 	\delta_2&=& \Delta_{21} -\Delta_{23}=  (q_2^a)^+- (q_2^a)^-  = -0.17121 , \\
 	\delta_3&=& \Delta_{31} +\Delta_{32}= (q_1^a)^+-(q_3^a)^-  = 0.46050. 
 \end{eqnarray}
 The two-particle phase shifts are computed to $\Delta_{12} =0.20572 $, $\Delta_{13} = 0.08358$ and $\Delta_{23} =0.37692 $ computed from formula (\ref{deltashift1}) with the values of the  asymptotic moment $p_i^-$ as stated in figure 2 for $g=1/4$.   The numerical values for the intersection of the asymptotes with the $y$-axis $(q_i^a)^\pm$  are reported in figure \ref{Calo4}.

 When $H=Q_3$ the ordering of the scattering is no longer controlled by the asymptotic momenta, but instead by the momenta squared.  Once more, we confirm the general formulas with the numerical values obtained from our example depicted in figure \ref{Calo4}
 \begin{eqnarray}
 	\delta_1&=& -\Delta_{12} -\Delta_{13}= (q_3^a)^+ - (q_1^a)^- =  -0.58264 , \\
 	\delta_2&=& \Delta_{21} -\Delta_{23}=  (q_2^a)^+- (q_2^a)^-  = 0.28929 , \\
 	\delta_3&=& \Delta_{31} +\Delta_{32}= (q_1^a)^+-(q_3^a)^-  = 0.29334,  
 \end{eqnarray}
 where $\Delta_{12} =0.37692$, $\Delta_{13} = 0.20572$ and $\Delta_{23} =0.08358 $ computed from formula (\ref{deltashift1}) with the values of the  asymptotic moment $p_i^-$ as stated in figure 2 for $g=1/4$. 
 
 \subsection{$A_6$- higher charge Calogero-Moser model}

Next, we demonstrate that the behaviour observed in the previous section is similar for higher rank $A_n$-Calogero theories, exemplified by the $A_6$-theory and  also reveal more features that are not observable in the $A_2$-theory with less particle in play.  

\begin{figure}[h]
	\centering         
	\begin{minipage}[b]{0.32\textwidth}          
		\includegraphics[width=\textwidth]{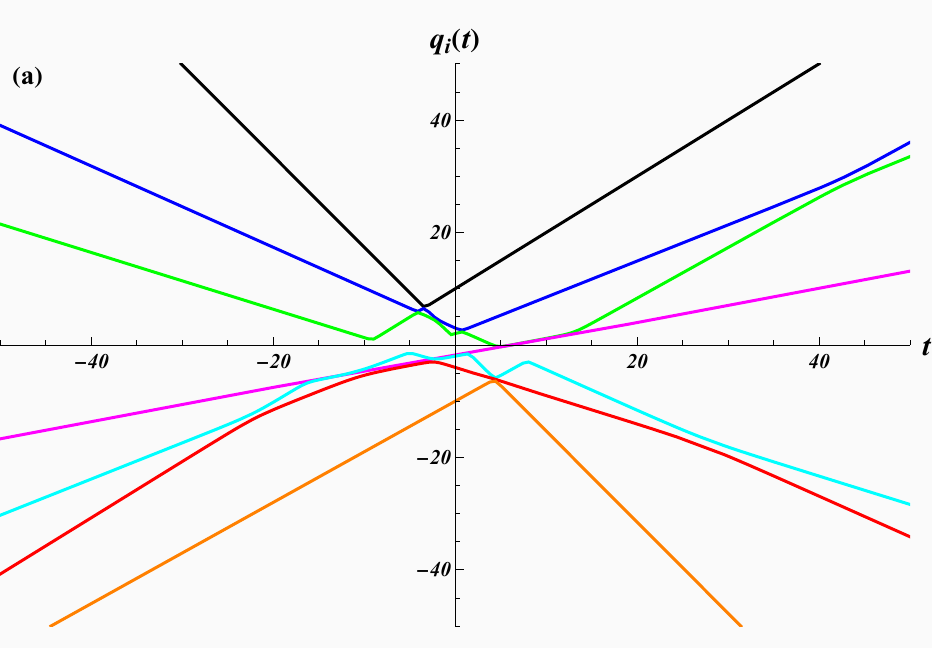}
	\end{minipage}   
	\begin{minipage}[b]{0.32\textwidth}           
		\includegraphics[width=\textwidth]{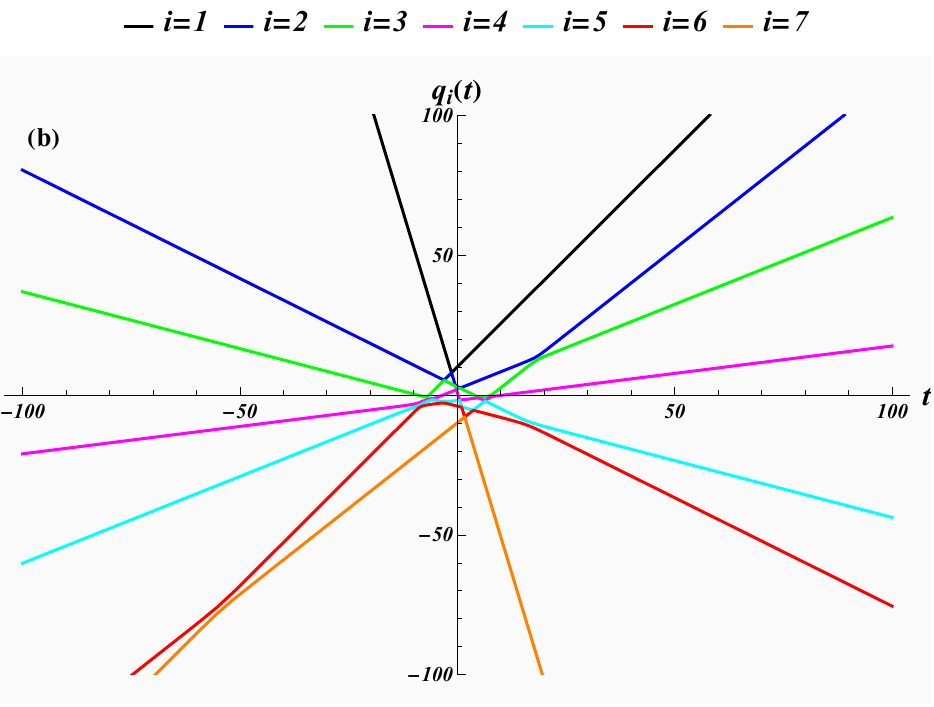}
	\end{minipage}  
	\begin{minipage}[b]{0.32\textwidth}           
		\includegraphics[width=\textwidth]{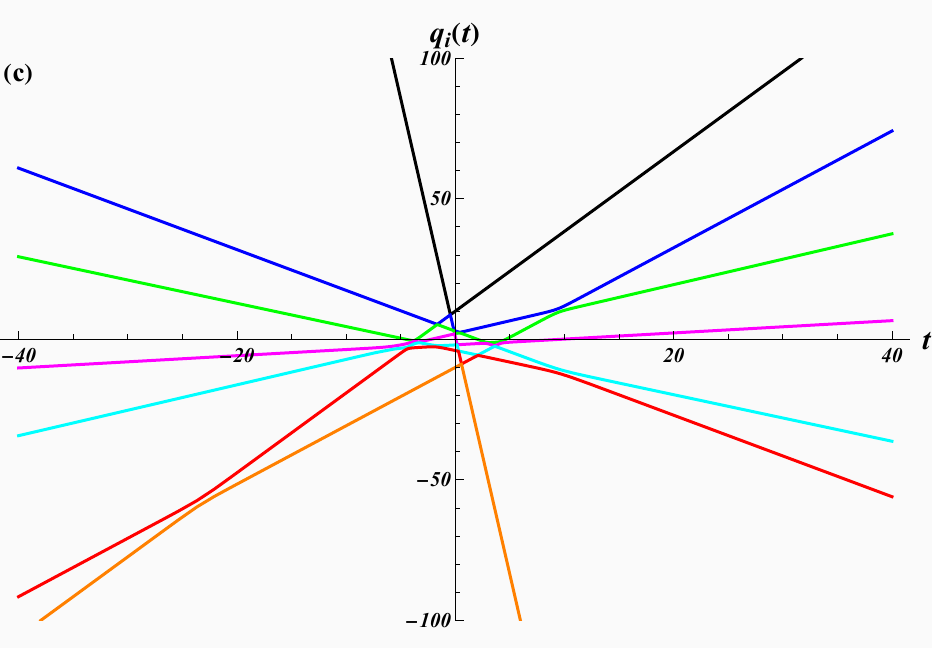}
	\end{minipage}  
	\begin{minipage}[b]{0.32\textwidth}          
		\includegraphics[width=\textwidth]{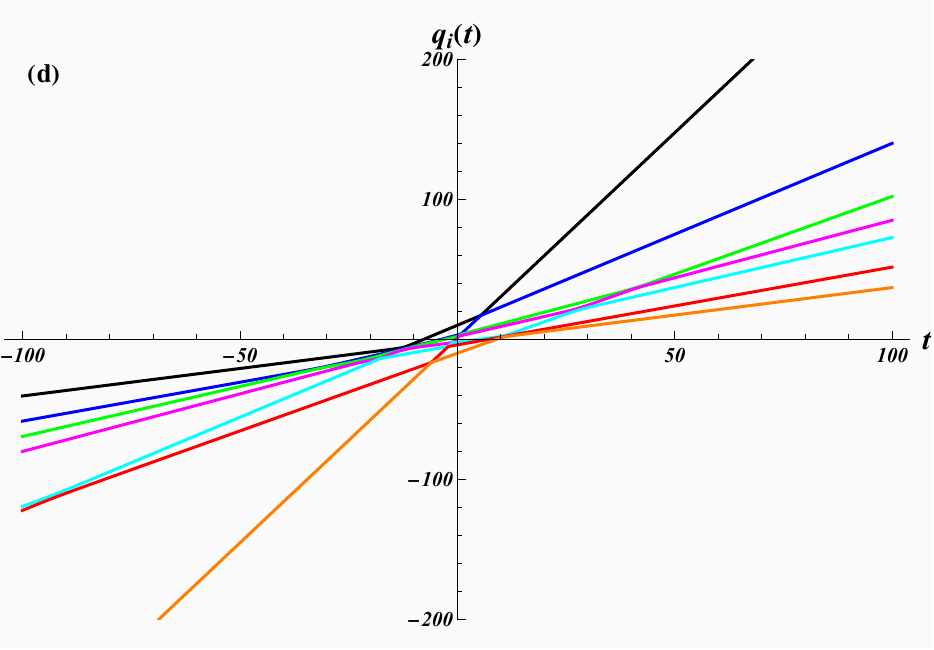}
	\end{minipage}   
	\begin{minipage}[b]{0.32\textwidth}           
		\includegraphics[width=\textwidth]{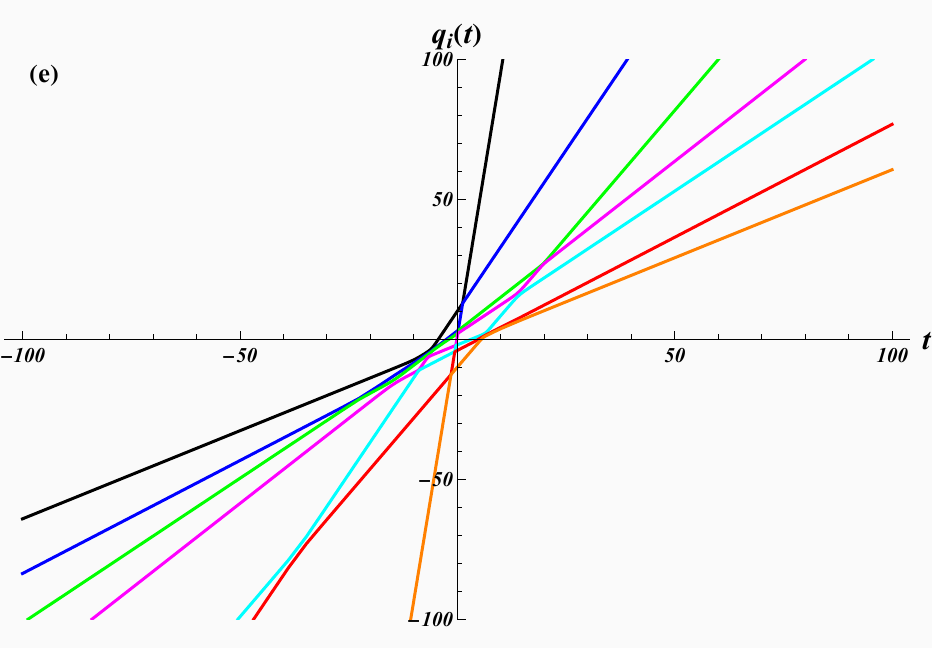}
	\end{minipage}  
	\begin{minipage}[b]{0.32\textwidth}           
		\includegraphics[width=\textwidth]{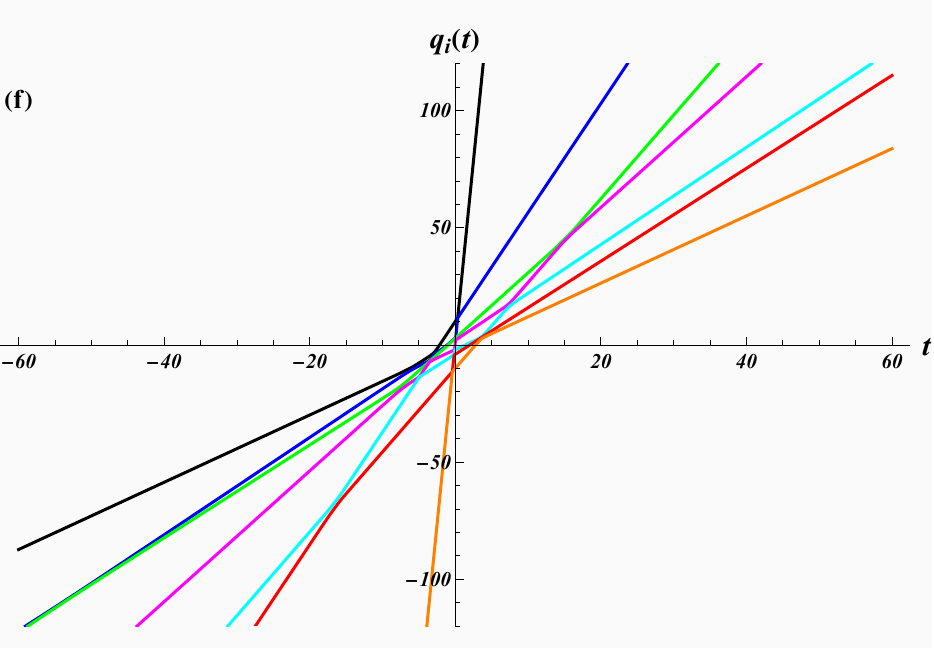}
	\end{minipage}  
	\caption{Coordinates $q_i(t)$ as functions of time $t$ for the $A_6$-Calogero model with $H=Q_2$ panel (a), $H=Q_4$ panel (b), $H=Q_6$ panel (c), $H=Q_3$ panel (d), $H=Q_5$ panel (e) and $H=Q_7$ panel (f). The initial conditions are taken as $q_1(0)=10$, $q_2(0)=3$, $q_3(0)=2$, $q_4(0)=1$, $q_5(0)=-2$, $q_6(0)=-4$, $q_7(0)=-10$, $p_1(0)=1$, $p_2(0)=-0.7$, $p_3(0)=0.6$, $p_4(0)=-1.6$, $p_5(0)=0.3$, $p_6(0)=-0.5$, $p_7(0)=0.9$ and the coupling constant as $g=0.05$.  } 
	\label{CaloA6coo}
\end{figure}
 
 \begin{figure}[h]
	\centering         
	\begin{minipage}[b]{0.32\textwidth}          
		  \includegraphics[width=\textwidth]{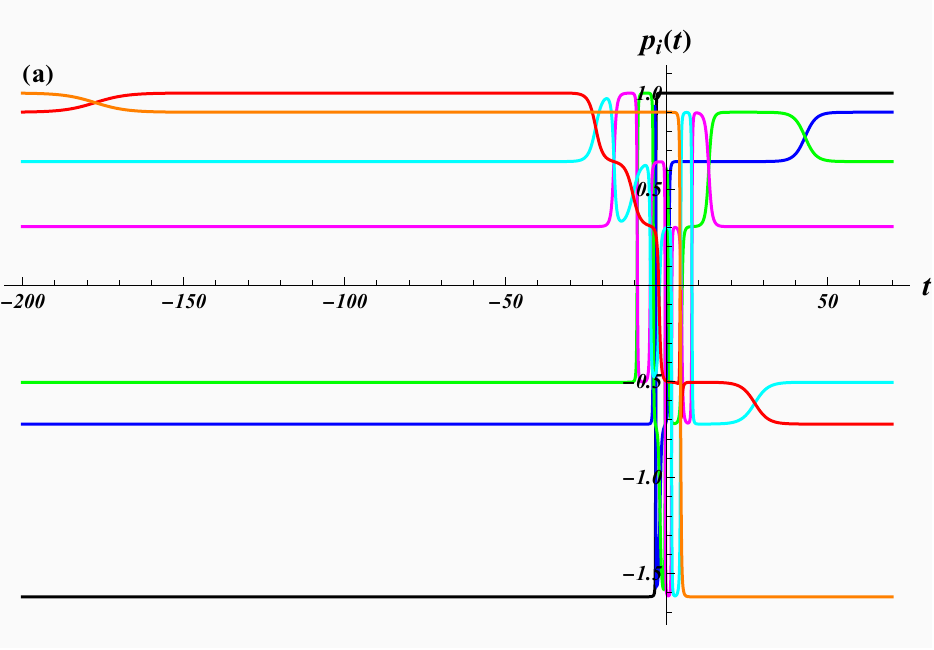}
	\end{minipage}   
	\begin{minipage}[b]{0.32\textwidth}           
	 	\includegraphics[width=\textwidth]{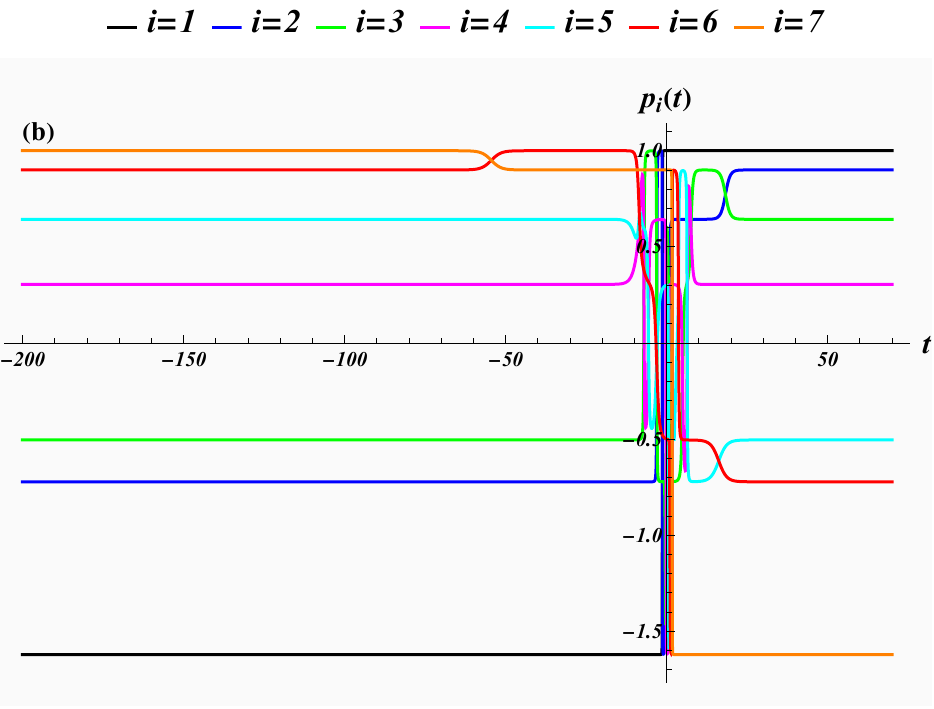}
	\end{minipage}  
	\begin{minipage}[b]{0.32\textwidth}           
		\includegraphics[width=\textwidth]{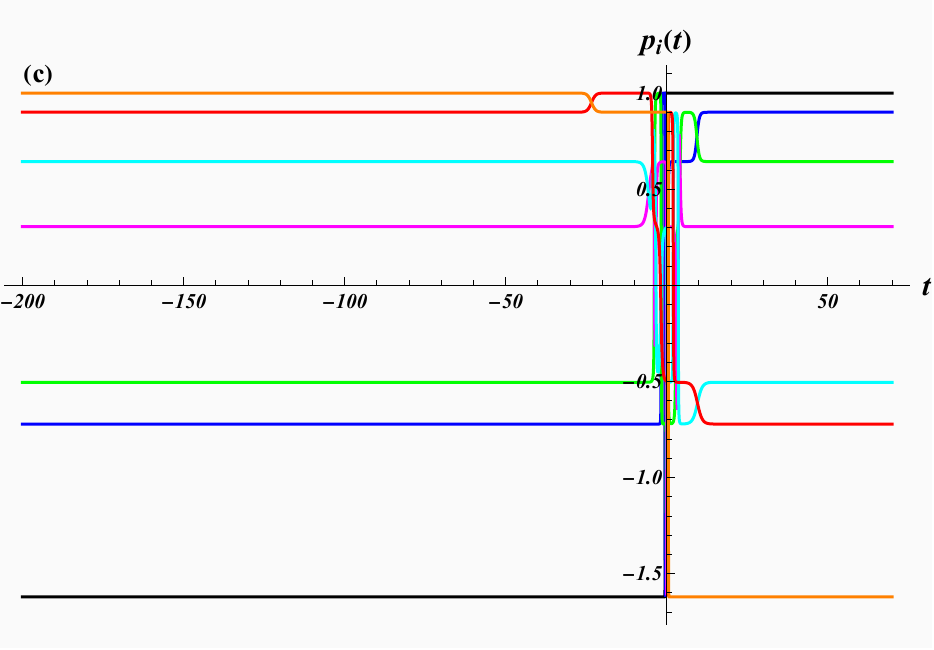}
	\end{minipage}  
	\begin{minipage}[b]{0.32\textwidth}          
		\includegraphics[width=\textwidth]{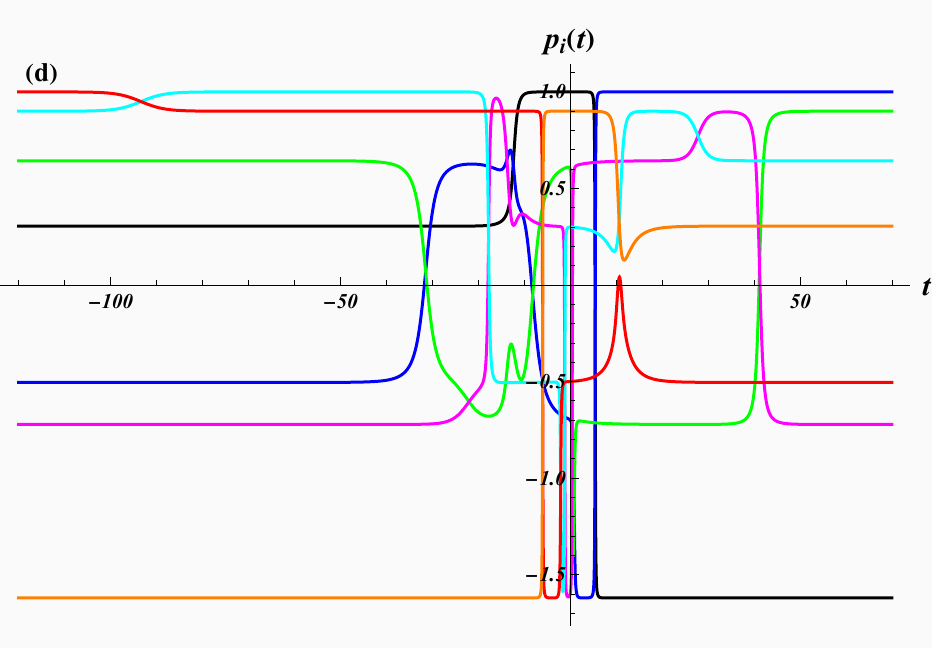}
	\end{minipage}   
	\begin{minipage}[b]{0.32\textwidth}           
		\includegraphics[width=\textwidth]{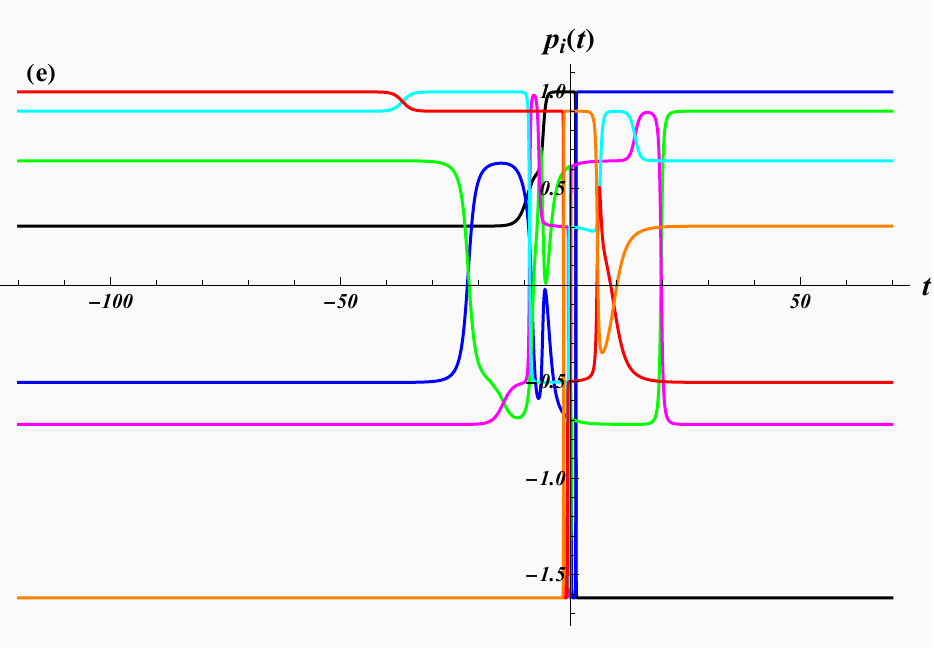}
	\end{minipage}  
	\begin{minipage}[b]{0.32\textwidth}           
		\includegraphics[width=\textwidth]{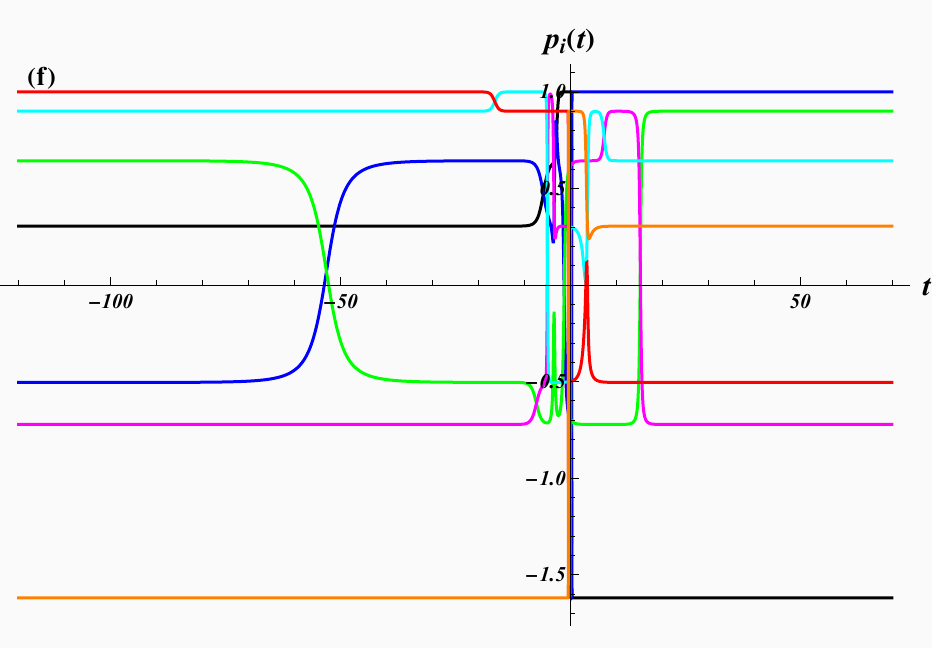}
	\end{minipage}  
	\caption{Momenta $p_i(t)$ as functions of time $t$ for the $A_6$-Calogero model with $H=Q_2$ panel (a), $H=Q_4$ panel (b), $H=Q_6$ panel (c), $H=Q_3$ panel (d), $H=Q_5$ panel (e) and $H=Q_7$ panel (f). The initial conditions and the coupling constant are the same as in figure \ref{CaloA6coo}. } 
	\label{CaloA6p}
\end{figure}

In the $q(t)$-plots in figure \ref{CaloA6coo} the most obvious feature that is confirmed is the fact that the spacial ordering does not change during the entire scattering process. Recalling that $\lim_{t \rightarrow \pm \infty} \dot{q}_i \sim (p_i^\pm)^{n-1} $ we also observe that $H=Q_m$ theories with $m$ odd can only have positive asymptotic gradients, whereas those with $m$ even have negative gradients for  $p_i^\pm <0$. As we argued in general at the beginning of section 3, the ordering of the particles is governed by the powers of the momenta as specified in (\ref{pordering}). This is confirmed by our observation in figure \ref{CaloA6p}. We see that in the $H=Q_2,Q_4,Q_6$-theories, panels (a)-(c), the asymptotic sets of momenta re-organise as
\begin{equation}
	p_1^- < 	p_2^- < 	p_3^- < 	p_4^- < 	p_5^- < 	p_6^- < 	p_7^-  \quad \rightarrow \quad
	p_7^+ < 	p_6^+ < 	p_5^+ < 	p_4^+ < 	p_3^+ < 	p_2^+ < 	p_1^+ ,   \label{peven}
\end{equation} 
so that the ordering on the line (\ref{qordering}) is enforced  by	$p_i^- < 	p_{i+1}^-$. In contrast, for the $H=Q_3,Q_5,Q_7$-theories,  panels (d)-(f),  the ordering is dictated by $(p_i^-)^2 < 	(p_{i+1}^-)^2$ with the momenta permuted as
\begin{equation}
\!\!	p_7^- < 	p_4^- < 	p_2^- < 	p_1^- < 	p_3^- < 	p_5^- < 	p_6^-  \quad \rightarrow \quad
	p_1^+ < 	p_4^+ < 	p_6^+ < 	p_7^+ < 	p_5^+ < 	p_3^+ < 	p_2^+.  \label{podd}
\end{equation} 
Using the six occurring asymptotic values of the momenta $p_1 = -1.621$, $p_2 = -0.722$, $p_3 = -0.505$, $p_4 = 0.305$, $p_5= 0.643$, $p_6 = 0.900$, $p_7 = 1.000$, we  calculate the two-particle phase shifts (\ref{deltashift1}) by suitably permuting them according to (\ref{peven}) and (\ref{podd}). We only report here three digits as this is sufficient for our arguments even though higher accuracy can of course be achieved. The values for $\Delta_{ij}$ are conveniently reported in a table:
\begin{center}
	\begin{tabular}{ |c||c|c|c|c|c|c|c|} 
		i / j & 1 & 2  & 3 & 4 & 5 & 6 & 7 \\  \hline  \hline
		1 & $\star$  &   0.111 & 0.074  & 0.026 & 0.019 & 0.016 & 0.014 \\ 
		2 & 0.111 &  $\star$ &   0.836     & 0.087    &  0.051     &  0.037      &  0.033    \\ 
		 3 & 0.074      &  0.836   &  $\star$    &   0.133    &  0.070     &  0.048     &  0.042    \\ 
		 4 &  0.026     &  0.087   &  0.133        &  $\star$   &   0.505    &  0.224     &   0.173   \\
		 5 &   0.019    &   0.051  &   0.070       &       0.505           &   $\star$     &  0.698     &   0.472    \\
		 6 &   0.016    &   0.037  &   0.048       &     0.224              &  0.698      &   $\star$     &  1.522    \\
		 7 &    0.014   &   0.033  &     0.042      &     0.173             &    0.472     &  1.522     &   $\star$   \\
	\end{tabular}
\end{center}
Using the table we compare next the two alternative ways to calculate the overall shift, i.e. directly from the asymptotic values and alternatively via (\ref{deltashift1}). The phase shifts for particle $i$ in the $H=Q_2,Q_4,Q_6$-theories are all identical resulting to 
\begin{eqnarray}
	\delta_1 &=&  -\Delta_{12}  -\Delta_{13} -\Delta_{14} -\Delta_{15} -\Delta_{16} -\Delta_{17}     =    (q^a_7)^+-(q^a_1)^- =-0.260 , \\
	\delta_2  &=&    \Delta_{21}  -\Delta_{23} -\Delta_{24} -\Delta_{25} -\Delta_{26} -\Delta_{27}     =(q^a_6)^+-(q^a_2)^-= -0.923 , \\
	\delta_3   &=&   \Delta_{31}  + \Delta_{32} -\Delta_{34} -\Delta_{35} -\Delta_{36} -\Delta_{37}    =(q^a_5)^+-(q^a_3)^-   =  0.606  , \\
	\delta_4   &=&    \Delta_{41}  + \Delta_{42} +\Delta_{43} -\Delta_{45} -\Delta_{46} -\Delta_{47}   =(q^a_4)^+-(q^a_4)^-= -0.656    ,   \\
	\delta_5    &=&    \Delta_{51}  + \Delta_{52} +\Delta_{53} +\Delta_{54} -\Delta_{56} -\Delta_{57}   =(q^a_3)^+-(q^a_5)^-  =-0.525 ,  \\
	\delta_6    &=&     \Delta_{61}  + \Delta_{62} +\Delta_{63} +\Delta_{64} +\Delta_{65} -\Delta_{67}   =(q^a_2)^+-(q^a_6)^-   = -0.500   ,    \\
	\delta_7    &=&       \Delta_{71}  + \Delta_{72} +\Delta_{73} +\Delta_{74} +\Delta_{75} +\Delta_{76}     =(q^a_1)^+-(q^a_7)^-    =  2.257    .   
\end{eqnarray}
Also the phase shifts for particle $i$ in the $H=Q_3,Q_5,Q_7$-theories are identical. Taking the permutation of the asymptotic momenta and the different orderings into account, we obtain
\begin{eqnarray}
	\delta_1 &=&    - \Delta_{43}  - \Delta_{45} - \Delta_{42} -\Delta_{46} -\Delta_{47} -\Delta_{41}               =(q^a_7)^+-(q^a_1)^-=-1.148,   \\
	\delta_2 &=&        \Delta_{34}  - \Delta_{35} - \Delta_{32} -\Delta_{36} -\Delta_{37} -\Delta_{31}              =(q^a_6)^+-(q^a_2)^-=-0.929, \\
	\delta_3&=&       \Delta_{54}  + \Delta_{53} - \Delta_{52} -\Delta_{56} -\Delta_{57} -\Delta_{51}           =(q^a_5)^+-(q^a_3)^-=-0.665,  \\
	\delta_4&=&        \Delta_{24}  + \Delta_{23} + \Delta_{25} -\Delta_{26} -\Delta_{27} -\Delta_{21}          =(q^a_4)^+-(q^a_4)^-=0.784,  \\
	\delta_5&=&           \Delta_{64}  + \Delta_{63} + \Delta_{65} +\Delta_{62} -\Delta_{67} -\Delta_{61}            =(q^a_3)^+-(q^a_5)^-=-0.531,   \\
	\delta_6&=&           \Delta_{74}  + \Delta_{73} + \Delta_{75} +\Delta_{72} +\Delta_{76} -\Delta_{71}         =(q^a_2)^+-(q^a_6)^-=2.228,   \\
	\delta_7&=&           \Delta_{14}  + \Delta_{13} + \Delta_{15} +\Delta_{12} +\Delta_{16} +\Delta_{17}         =(q^a_1)^+-(q^a_7)^-=0.260.   
\end{eqnarray}
We also verify that in all cases the total overall shift is zero, $\sum_{i=1}^7  \delta_i =0 $. While the $H=Q_m$-theories share a lot of features for even and odd $m$, there are also a number of differences. We notice, especially from figure \ref{CaloA6p}, that for large $m$ the actual scattering event becomes more and more squeezed around $t=0$. The reason for this is that the singularities in potentials acquire higher order and, as a result, become more short-ranged. Another noteworthy feature is the change in the ordering in which the particles scatter with respect to the value of $m$. At present, we lack an explanation that could predict these different scattering sequences.

\section{A comment on the quantum mechanical scattering}
We finish with a brief comment on the quantum theory. In general, we will have to rely on numerical solutions for the higher particle theories. However, for the $A_2$-Calogero model we can extend the known analytic solution of the time-independent Schr\"odinger equation \cite{Cal2,marchiorosol} to the $H=Q_3$-theory. 
For this purpose we first transform our system to the centre-of-mass coordinates using Jacobi coordinates\footnote{When changing the factor $1/3$ in the definition of $R$ to $1/\sqrt{3}$ the new coordinates simply result from taking a two dimensional representation for the $A_2$ simple roots, see for instance \cite{bethanAF}.}
\begin{equation}
	R=\frac{1}{3}(q_1+q_2+q_3), \quad \zeta=\frac{1}{\sqrt{6}}(q_1+q_2-2q_3 ),\quad
	\eta =\frac{1}{\sqrt{2}}(q_1-q_2).
\end{equation} 
Setting $R$ to zero the remaining two dimensional system is further transformed to the polar coordinates $r$ and $\phi$ with
\begin{equation}
	\zeta= r \cos(\phi), \qquad  \eta= r \sin(\phi).
\end{equation} 
In these variable the second and third charge, (\ref{Q2charge}) and (\ref{Q3charge}) take on the form
\begin{eqnarray}
	Q_2 &=& \frac{1}{2} \left(  - \partial_{rr} - \frac{1}{r}  \partial_{r}   - \frac{1}{r^2}  \partial_{\phi \phi}    + \frac{9 g}{r^2 \sin^2 (3 \phi)}  \right) ,\\
	Q_3 &=& \frac{i \sin (3 \phi )}{3 \sqrt{3} r^3}  \left\{    \left(9 g \csc ^2(3 \phi )+16\right) \partial_\phi  -2 \partial_{\phi\phi\phi}-12 \cot (3 \phi ) \partial_{\phi\phi}   +  r \left[ 6 r \partial_{r r\phi} -18  \partial_{r \phi}  \right.  \right. \\
	&& 
	\left. \left.   -3 \cot (3 \phi ) \left(9 g \csc
	^2(3 \phi )+2\right) \partial_{r } +6 \cot (3
	\phi ) \partial_{r \phi \phi} 
	-2 r \cot (3 \phi ) \left(r \partial_{rrr }-3 \partial_{rr}
	\right)\right]  \right\} .   \qquad \quad \notag
\end{eqnarray}
We convince ourselves that these charges commute, i.e. $[Q_2,Q_3]=0$. The most general  eigenfunction $\psi$ to $Q_2 \psi = E \psi$ was found in \cite{Cal2} as
\begin{equation} 
	\psi(r,\phi,g,E)  = \sum_{\ell=0}^\infty  \psi_\ell(r,\phi,g,E)=\sum_{\ell=0}^\infty  N_\ell J_{3 a+3
		\ell +\frac{3}{2}} \left( \sqrt{E} r\right)  C_\ell^{a+\frac{1}{2} }   \left[ \cos (3 \phi ) \right]   \sin ^{a+\frac{1}{2}}(3 \phi ), \label{solpsi}
\end{equation} 
where $J_n(z)$ denotes the Bessel function of the first kind, $C_n^m$ a Gegenbauer polynomial, $N_\ell $ is a normalisation constant and $a:=\frac{1}{2} \sqrt{2 g+1}$. This is indeed a scattering state for the continuous eigenvalue $E$. Since $Q_2$ and $Q_3$ commute also $Q_3 \psi$ must be an eigenstate of $Q_2$. Indeed, we find 
\begin{equation} 
	Q_3 \psi=  \frac{i E^{3/2}}{3 \sqrt{3} }  \left[ \sum_{\ell=0}^\infty     a_\ell  \psi_{\ell+1} +\sum_{\ell=1}^\infty   b_\ell \psi_{\ell-1}\right]  =  \frac{i E^{3/2}}{3 \sqrt{3} }  \left[ \sum_{\ell=0}^\infty  \left(    a_{\ell-1} + b_{\ell+1} \right)  \psi_\ell     \right]  ,
\end{equation} 
with 
\begin{equation} 
	a_\ell  = (\ell+1) \frac{2 \ell +1 -\sqrt{2 g+1} }{2 \ell (\ell+1)-g }, \quad b_\ell = a_\ell- 2. 
\end{equation} 
The expressions can be used to extract quantum scattering matrices from the asymptotic behaviour. We leave these investigation for a separate analysis \cite{bethanAFprep}. 

\section{Conclusion}
We generalised the standard integrable scattering theories based on conventional Hamiltonians quadratic in the momenta to theories involving higher powers in momenta and additional non-potential terms. As candidates for these theories we used higher order charges of integrable systems. We discussed as particular examples the scattering behaviour of Calogero and Calogero-Moser systems related to the $A_2$ and $A_6$ Lie algebras. We derived the overall and two-particle classical scattering phase shifts. Based on the fact that all higher charge theories allow for a Lax pair formulation with an $L$-operator in common with the Hamiltonian system, we argued that they possess the same set of asymptotic momenta. Moreover, these systems also possess the trademark property of integrable systems of their overall shift being the sum of all two particle scattering events. However, due to the different asymptotic dependencies of the velocities on the momenta, the summation differs for different $m$ in the $H=Q_m$-theories.

Naturally there are a  number of interesting open questions left to be answered in future work. At present we can precisely which two-particle scattering processes take place in dependence on $m$, but not the precise order in which they take place. It would be interesting to find out how these results generalise to Calogero and Calogero-Moser models based on other type of algebras, representations of the root systems, to integrable systems of different type and to non-integrable systems in general. More detailed analysis on the non-integrable systems in form of perturbed integrable systems would reveal more information about the robustness of the features obtained here. Of further interest is also a more detailed analysis of the quantum version of higher derivative theories \cite{bethanAFprep}.

\medskip
\noindent \textbf{Acknowledgments:} BT is supported by a City, University of London Research Fellowship.

\newif\ifabfull\abfulltrue

%%\bibliographystyle{phreport}
%%\bibliography{acompat,Ref}

\end{document}